\documentclass{article}
\usepackage{amsmath,amsthm}
\usepackage{amssymb,latexsym}
\usepackage[mathscr]{eucal}
\usepackage{setspace}
\usepackage{graphics}
\usepackage{array}
\usepackage{nccmath}
\usepackage{geometry}
\usepackage{lineno}
\usepackage[affil-it]{authblk}
\newcommand{\tbf}[1]{\textbf{#1}}
\newcommand{\tit}[1]{\textit{#1}}
\newcommand{\del}[0]{\partial}
\newcommand{\phihat}[0]{\hat{\phi}}

\begin{document}
\newgeometry{top = 2in}

\title{\textbf{Potential Flow Theory Formulation of Parker's Unsteady Solar Wind Model and Nonlinear Stability of Parker's Steady Solar Wind Solution}}
\author{Bhimsen K. Shivamoggi$^{1,2}$\footnote{}}
\affil{$^1$California Institute of Technology \\ Pasadena, CA 91125, USA}
\affil{\textnormal{$^2$Permanent Address: University of Central Florida} \\  \textnormal{Orlando, FL, 32816, USA}}
\date{}
\maketitle

\begin{abstract}The purpose of this paper is to present a novel optimal theoretical framework based on \tit{potential flow} theory in ideal gas dynamics which provides a smooth extrapolation of Parker's steady solar wind model to the unsteady case. The viability of this framework is illustrated by providing the first ever systematic theoretical formulation to successfully address the long-standing open issue of regularization of the singularity associated with the Parker sonic critical point (where the solar wind flow velocity equals the speed of sound in the gas) in the linear stability problem of Parker's steady solar wind solution. This development involves going outside the framework of the linear perturbation problem and incorporating the dominant nonlinearities in this dynamical system, and hence provides an appropriate nonlinear recipe to regularize this singularity. The stability of Parker's steady wind solution is shown to extend also to the neighborhood of the Parker sonic critical point by analyzing the concomitant nonlinear problem. The new theoretical framework given here seems, therefore, to have the potential to provide a viable basis for future formulations addressing various theoretical aspects of Parker's unsteady solar wind model.
\end{abstract}


\vspace{\fill}
\noindent Corresponding Author: Bhimsen K. Shivamoggi\\
bhimsen.shivamoggi@ucf.edu

\restoregeometry
\newpage

\section{Introduction} The solar wind is a hot tenuous magnetized plasma outflowing continually from the sun, which carries off a huge amount of angular momentum from the sun while causing only a negligible mass loss. The bulk of the solar wind is known to emerge from coronal holes (Sakao et al. 2007) and to fill the heliosphere (Dialynas et al. 2017). Recent Parker Solar Probe\footnote{The Parker Solar Probe (Shivamoggi 2024) has been providing significant information on the conditions in the inner solar corona (Fisk and Casper 2020, Bowen et al. 2022, and others) some of which were at variance with previous belief (like the coupling of the solar wind with solar rotation (Kasper et al. 2019), which was shown (Shivamoggi 2020) to cause enhanced angular momentum loss from the sun).} observations (Bale et al. 2023) indicated that the \tit{fast} solar wind emerges from the coronal holes via the process of magnetic reconnection between the open and closed magnetic field lines (called the \tit{interchange reconnection}). On the other hand, recent \tit{in situ} observations from Solar Orbiter (Yardley et al. 2024) indicated the latter process to be the cause of the \tit{slow} solar wind as well. Coronal heating along with high thermal conduction is believed to be the cause of weak to moderate-speed solar wind. But some additional acceleration mechanism operating beyond the coronal base seems to be needed for high-speed solar wind (Parker 1958, 1965). Parker (1958) gave an ingenious model to accomplish this by continually converting the coronal thermal energy into kinetic energy of the wind and accelerating the latter from subsonic to supersonic speeds\footnote{Recent \emph{in situ} measurements (Rivera et al. 2024) of a given plasma patch (identified by an Alfv\'enic magnetic field line switchback) from two solar space probes (Parker Solar Probe at 13.3 solar radii and Solar Orbiter at 127.7 solar radii) revealed an enormous drop in the magnetic activity associated with this patch as it moved past these space probes along with a major heating and acceleration experienced by the patch.}. The various physical properties in the solar wind have been confirmed by \textit{in situ} observations (Meyer-Vernet 2007). 

Parker's steady-solar wind solution has the significance of,
\begin{itemize}
\item{being a physically acceptable solution that describes a smooth acceleration of the solar wind through the sonic conditions at the \textit{Parker sonic critical point}, given by $r = r_*= GM_S/2a^2$, $G$ is the gravitational constant, $M_S$ is the mass of the sun, and $a$ is the speed of sound in the wind;}
\item{satisfying a special boundary condition prescribing the pressure to decrease away from the sun to zero at infinity in the interstellar space.}
\end{itemize}
On the other hand, solar wind observations (Schrijver 2000) indicated that the large-scale behavior of the solar wind, on the average, its local noisiness (Feldman et al. 1977) notwithstanding, is apparently close to Parker's solar wind solution. This indicates that Parker's solar wind solution apparently possesses a certain robustness and an ability to sustain itself against small perturbations acting on this system. Parker (1960) therefore proposed that his solution exhibits an intrinsic stability like a "\textit{stable attractor}" of this dynamical system (Cranmer and Winebarger 2019). So, any deviations in flow variables from Parker's solar wind solution, Parker (1960) argued, would be convected out by the wind flow and damped out.

This renders the stability of Parker's solar wind solution an important issue, though still not completely resolved. This issue was investigated by Parker (1966) via formal considerations of the dynamical equations governing the solar wind flow. Parker (1966) advocated that the stability of the flow in the \textit{subcritical} region inside the Parker sonic critical point may be investigated by approximating the solar corona in this region by a static atmosphere on the grounds that no intrinsic shear-flow instabilities may be generated in the corona during its expansion in this region\footnote{This is compatible with the absence of coronal-flow shear in the spherically symmetric flow situation posited in Parker's solar wind model (Parker 1958), which would otherwise become a free-energy source of these shear-flow instabilities (Shivamoggi 1986).}. Shivamoggi (2023) followed up on Parker's proposition for the subcritical region, and gave a systematic analytical development of this issue, by posing a Sturm-Liouville problem for the linearized perturbations about Parker's steady solar wind solution, to demonstrate its intrinsic stability.

On the other hand, the investigation of stability of Parker's solar wind solution with respect to linearized perturbations with the inclusion of solar wind flow in the basic state was initiated by Parker (1966), Carovillano and King (1966), and Jockers (1968). They found that the linear perturbation problem possesses a singularity at the Parker sonic critical point which makes this problem ill-posed\footnote{Later Velli (1994) tried, apparently unaware of these developments, to extrapolate the linear perturbation problem to cover the ill-posed linear transonic flow region around the Parker sonic critical point. Consequently, the veracity of Velli's (1994) results is open to question.}. This leads to preclusion of well-behaved solutions of the linear perturbation problem in the \textit{transonic} flow region (where the wind flow-speed is near the speed of sound in the gas) around the Parker sonic critical point.

We wish to point out that a regularization of this singularity requires going outside of the framework of the linear perturbation problem and incorporating the dominant nonlinearities in this dynamical system (akin to the situation in \tit{transonic aerodynamics} (Dowell 2022, Shivamoggi 2023)). The straightforward unsteady version of Parker's solar wind model used in Parker (1966), Carovillano and King (1966), Jockers (1968), Shivamoggi (2023) for stability considerations lends a rather cumbersome mathematical approach toward this objective.  The purpose of this paper is to present a whole new theoretical formulation of Parker's unsteady solar wind model based on the \textit{potential flow theory} in ideal gas dynamics, which provides an optimal theoretical framework to analyze various theoretical aspects of Parker's unsteady solar wind model in general, and regularization of the singularity at the Parker sonic critical point in the linear stability problem in particular, by going to the concomitant nonlinear problem.\\

\section{Potential-Flow Formulation of Parker's Unsteady Solar Wind Model.} 

Consider an ideal gas flow in the presence of a central gravitating point mass representing the sun. The solar wind is represented by a spherically symmetric flow so the flow variables depend only on the distance $r$ from the sun and time $t$, and the flow velocity is taken to be only in the radial direction.

The equations expressing the conservation of mass and momentum balance for the ideal gas flow constituting the solar wind are (in usual notations),
\begin{equation}
\frac{\partial \rho}{\partial t} + \frac{1}{r^2} \frac{\partial}{\partial r} \left( \rho r^2 v \right) = 0
\label{eq::1}
\end{equation}

\begin{equation}
\rho \left( \frac{\partial v}{\partial t} + v \frac{\partial v}{\partial r} \right) = - \frac{\partial p}{\partial r} - \rho \frac{d U}{dr}
\label{eq::2}
\end{equation}
where $U$ is the gravitational potential associated with the sun,
\begin{equation}
U = -\frac{G M_S}{r}
\label{eq::3}
\end{equation}
We assume the ideal gas flow under consideration to be modeled by a \textit{potential flow}, so we introduce a velocity potential $\Phi$, according to
\begin{equation}
v = \frac{\partial \Phi}{\partial r}
\label{eq::4}
\end{equation}
Furthermore, we assume, for analytical simplicity, that the gas flow occurs under \textit{isothermal} conditions, so
\begin{equation}
p = a^2 \rho
\label{eq::5}
\end{equation}
where $a$ is the constant speed of sound in the gas\footnote{SOHO observations (Cho et al. 2018) indicated that the solar wind expands isothermally to considerable distances.}. In the same vein, we assume that the flow variables as well as their derivatives vary continuously so there are no shocks occurring anywhere in the region under consideration.

Using (\ref{eq::4}), equations (\ref{eq::1}) and (\ref{eq::2}) become
\begin{equation}
\frac{1}{\rho} \frac{\partial \rho}{\partial t} + \left( \frac{\partial^2 \Phi}{\partial r^2}+ \frac{2}{r} \frac{\partial \Phi}{\partial r} \right) + \frac{1}{\rho} \frac{\partial \Phi}{\partial r} \frac{\partial \rho}{\partial r} = 0
\label{eq::6}
\end{equation}
\begin{equation}
\frac{\del}{\del t} \left( \frac{\del \Phi}{\del r} \right) + \frac{\del \Phi}{\del r} \frac{\del^2 \Phi}{\del r^2} = -\frac{1}{\rho} \frac{\del p}{\del r} - \frac{d U}{dr}.
\label{eq::7}
\end{equation}

The Bernoulli integral of equation (\ref{eq::7}),
\begin{equation}
\frac{\del \Phi}{\del t} + \frac{1}{2} \left( \frac{\del \Phi}{\del r} \right)^2 + \int \frac{dp}{\rho} + U = const
\label{eq::9}
\end{equation}
on using (\ref{eq::5}), gives,
\begin{equation}
\frac{1}{\rho} \frac{\del \rho}{\del t} = -\frac{1}{a^2} \left( \frac{\del^2 \Phi}{\del t^2} + \frac{\del \Phi}{\del r} \frac{\del^2 \Phi}{\del t \del r} \right)
\label{eq::10}
\end{equation}
Using equations (\ref{eq::7}) and (\ref{eq::10}), equation (\ref{eq::6}) leads to the equation governing the potential flows of an ideal gas constituting the solar wind,
\begin{equation}
\left[ a^2 - \left( \frac{\del \Phi}{\del r} \right)^2 \right] \frac{\del^2 \Phi}{\del r^2} + \frac{2 a^2}{r} \frac{\del \Phi}{\del r} = \frac{\del^2 \Phi}{\del t^2} + 2 \frac{\del \Phi}{\del r} \frac{\del^2 \Phi}{\del t \del r} + \frac{\del \Phi}{\del r} \frac{dU}{dr}.
\label{eq::11}
\end{equation}

Equation (\ref{eq::11}) provides an optimal theoretical framework to extrapolate Parker's solar wind model to unsteady situations in general, and investigate in particular, the issue of stability of the Parker steady solar wind solution in a compatible manner.\\

\section{Parker Steady Solar Wind Model.}

For a steady solar wind flow, using (\ref{eq::3}), equation (\ref{eq::11}) describes Parker's solar wind model (Parker 1958),
\begin{equation}
\left[ a^2 - \left( \frac{d \Phi}{dr} \right)^2 \right] \frac{d^2 \Phi}{dr^2} + \frac{2 a^2}{r^2}\left( r - r_* \right) \frac{d \Phi}{dr} = 0.
\label{eq::12}
\end{equation}

Equation (\ref{eq::12}) gives the physically acceptable smooth solution (Parker 1958),
\begin{equation}
\left[ \frac{d \Phi/dr}{a} \right]^2 - \log\left[ \frac{d\Phi/dr}{a} \right]^2 = 4 \log\left( \frac{r}{r_*} \right) + 4\left( \frac{r}{r_*}\right)-3
\label{eq::13}
\end{equation}
which complies with the smoothness condition at the Parker sonic critical point,
\begin{equation}
r = r_*: \hspace{.1in} \frac{d \Phi}{dr}=a.
\label{eq::14}
\end{equation}\\\

\section{Linear Perturbation Problem for Parker's Solar Wind Model.}

We consider solutions of time-dependent perturbations (denoted by subscript $1$) on the basic state (denoted by subscript $0$), to be of the form,
\begin{equation}
\Phi(r,t) = \phi_0(r)+ \epsilon \phi_1(r,t), \hspace{.1in} \epsilon \ll 1
\label{eq::15}
\end{equation}
and assume the perturbations characterized by the small parameter $\epsilon$ to be small. Equation (\ref{eq::11}) then yields for the basic state,
\begin{equation}
\left[ a^2 - \left( \frac{d \phi_0}{dr} \right)^2 \right] \frac{d^2 \phi_0}{dr^2} + \frac{2 a^2}{r^2}(r-r_*) \frac{d \phi_0}{dr} = 0
\label{eq::16}
\end{equation}
which represents Parker's steady solar wind model given by equation (\ref{eq::12}), and the following equation for the linearized perturbations,
\begin{equation}
\left[ a^2 - \left( \frac{d \phi_0}{dr} \right)^2 \right] \frac{\del^2 \phi_1}{\del r^2} + \left[-2 \frac{d \phi_0}{dr} \frac{d^2 \phi_0}{dr^2} +2a^2 (r-r_*) \right] \frac{\del \phi_1}{\del r} = 2 \frac{d \phi_0}{dr} \frac{\del^2 \phi_1}{\del t \del r} + \frac{\del^2 \phi_1}{\del t^2}.
\label{eq::17}
\end{equation}

We consider the subcritical region, where
\begin{equation}
\left[ a^2 - \left( \frac{d \phi_0}{dr} \right)^2 \right] >0, \hspace{.1in} r < r_*
\label{eq::18}
\end{equation}
and assume normal-mode solutions of the form,
\begin{equation}
\phi_1(r,t) = \hat{\phi}_1(r)e^{-i \omega t}.
\label{eq::19}
\end{equation}
Equation (\ref{eq::17}) then gives
\begin{multline}
\frac{d^2 \hat{\phi}_1}{dr^2} + \left[ -\frac{d^2 \phi_0/dr^2}{d\phi_0/dr} - 2 \frac{(d\phi_0/dr) (d^2\phi_0/dr^2)}{a^2 - (d \phi_0/dr)^2} + 2i \omega \frac{d\phi_0/dr}{a^2 - (d\phi_0/dr)^2} \right] \frac{d \phihat_1}{dr}\\ + \omega^2 \left[ \frac{1}{a^2-(d \phi_0/dr)^2}\right] \phi_1 = 0.
\label{eq::20}
\end{multline}

Equation (\ref{eq::20}) may be written as the \textit{Sturm-Liouville} equation (Birkhoff and Rota 1989),
\begin{equation}
\frac{d}{dr}\left[ f(r) \frac{d \phihat_1}{dr} \right] + \omega^2 g(r) \phihat_1 = 0, \hspace{.1in} r_S < r < r_*
\label{eq::21}
\end{equation}
where
$$
f(r) \equiv \left[ \frac{a^2- (d\phi_0/dr)^2}{d\phi_0/dr} \right] e^{2i\omega \int_{r_S}^r \frac{d\phi_0/dr}{a^2-(d\phi_0 dr)^2}dr}
$$
$$
g(r) \equiv \frac{1}{d\phi_0/dr}e^{2i\omega \int_{r_S}^r \frac{d\phi_0/dr}{a^2-(d\phi_0/dr)^2}dr}
$$
$r_S$ being sun's radius.

If we assume $\omega$ to be purely imaginary, $\omega = i \Omega$, equation (\ref{eq::21}) becomes
\begin{equation}
    \tag{21a}
    \frac{d}{dr} \left[ \tilde{f}(r) \frac{d \hat{\phi}_1}{dr} \right] - \Omega^2 \tilde{g}(r) \hat{\phi} = 0, \hspace{0.1in} r_s < r < r_*
    \label{eq::21a}
\end{equation}
where,
$$
\tilde{f}(r) = \left[ \frac{a^2- (d\phi_0/dr)^2}{d\phi_0/dr} \right] e^{-2\Omega \int_{r_S}^r \frac{d\phi_0/dr}{a^2-(d\phi_0/dr)^2}dr} > 0
$$
$$
\tilde{g}(r) = \frac{1}{d\phi_0/dr}e^{-2\Omega \int_{r_S}^r \frac{d\phi_0/dr}{a^2-(d\phi_0/dr)^2}dr} > 0.
$$
Taking the complex conjugate of equation (\ref{eq::21a}), we have
\begin{equation}
    \tag{21b}
    \frac{d}{dr} \left[ \tilde{f}(r) \frac{d \bar{\hat{\phi}}_1}{dr} \right] - \Omega^2 \tilde{g}(r) \bar{\hat{\phi}}_1 = 0, \hspace{.1in} r_s < r < r_*
    \label{eq::21b}
\end{equation}
We then obtain from equations (\ref{eq::21a}) and (\ref{eq::21b}),
\begin{equation}
    \setcounter{equation}{22}
-\int_{r_S}^r \tilde{f}(r) \left| \frac{d \phihat_1}{dr} \right|^2 dr - \Omega^2 \int_{r_S}^r \tilde{g}(r) \left| \phihat_1 \right|^2 dr = 0, \hspace{.1in} r_S< r< r_*
\label{eq::23}
\end{equation}
where we have taken the perturbations or their gradients to vanish at the coronal base $r = r_S$. Equation (\ref{eq::23}) is impossible to satisfy, so $\omega$ is real\footnote{In non-dissipative systems (like the one under consideration) the transition from stability to instability may be expected to occur via a marginal state exhibiting oscillatory motions (Eddington 1926, see also Chandrasekhar 1961).}, and Parker's solar wind solution is linearly stable in the subcritical region.

It is to be noted, as previously mentioned by Parker (1966), Carovillano and King (1966) and Jockers (1968), that the linearized perturbation problem, described by equation (\ref{eq::20}), exhibits a singularity at the Parker sonic critical point given by (\ref{eq::14})\footnote{It may be mentioned, as Parker (1966) pointed out, that this coincidence will not hold for more general non-isothermal cases. On the other hand, in the latter case, the variations in the sound speed would lead to additional nonlinearities, which can also materialize in regularizing the singularity in the linear pertubation probelm.}. Consequently, the above linearized development, which is valid in the subcritical region, becomes ill-posed and breaks down near the Parker sonic critical point. This drawback may be remedied via a proper treatment of the transonic flow region around the Parker sonic critical point, which necessitates going outside the linearized framework given above, and adopting the nonlinear formulation (akin to the situation in \textit{transonic aerodynamics} (Dowell 2022, Shivamoggi 2023)). This task can be accomplished in an expeditious way by using the potential-flow theory formulation, namely equation (\ref{eq::11}), given in this paper.\\

\section{Nonlinear Perturbation Problem for Parker's Solar Wind Model.}

Equation (\ref{eq::11}) governing the potential flows of an ideal gas constituting the solar wind may be rewritten as,
\begin{equation}
\left[ a^2 - \left( \frac{\del \Phi}{\del r} \right)^2 \right] \frac{\del^2 \Phi}{\del r^2}+ \frac{2 a^2}{r^2}(r-r_*) \frac{\del \Phi}{\del r} = \frac{\del^2 \Phi}{\del t^2} + 2 \frac{\del \Phi}{\del r}\frac{\del^2 \Phi}{\del t \del r}
\label{eq::24}
\end{equation}
which extends equation (\ref{eq::12}) to the unsteady cases. The singularity at the Parker sonic critical point, given by $\del \Phi/\del r = a$, indicated by equation (\ref{eq::24}), renders this problem a \tit{singular perturbation problem}. In order to resolve this singularity, we follow the \tit{thin airfoil theory in transonic flows} (Cole and Messiter 1957), and look for a solution for the perturbations, valid near the Parker sonic ciritical point, following \tit{method of multiple scales} (Shivamoggi 2003), of the form,

\begin{equation}
\begin{gathered}
\frac{\del \Phi}{\del r} = a \left( 1 + \epsilon \frac{\del \phi_1}{\del r} \right),  \\
r=r_*(1+\epsilon x),  \hspace{.1in} \tilde{t} =\epsilon t, \hspace{.1in} \epsilon \ll 1
\end{gathered}
\label{eq::25}
\end{equation}
where the small parameter $\epsilon$ (introduced previously in (\ref{eq::15}) to characterize the pertubations), for analytical simplicity, is also taken to characterize the deviation of the flow speed from the speed of sound in the wind. The slow time scale $\tilde{t}$ characterizes the \tit{slowly-varying dynamics} influenced by gravitational choking (see Shivamoggi 2020) operational near the Parker sonic critical point. Equation (\ref{eq::24}) then yields, on keeping the dominant terms,

\begin{equation}
    \label{eq::25'}
    \frac{\del^2 \phi_1}{\del \tilde{t} \del r} + a \frac{\del \phi_1}{\del r} \frac{\del^2 \phi_1}{\del r^2} = \frac{a}{r^*} x \frac{\del \phi_1}{\del r}.
\end{equation}
Introducing the normalized radial flow-velocity perturbation,
\begin{equation}
    \label{eq::26}
    u_1 \equiv \frac{\del \phi_1}{\del r}
\end{equation}
equation (\ref{eq::25'}) becomes
\begin{equation}
    \label{eq::27}
    \frac{\del u_1}{\del \tilde{t}} + a u_1 \frac{\del u_1}{\del r} = \left( \frac{x}{\tau} \right) u_1.
\end{equation}
where $\tau \equiv r_*/a$ may be interpreted as the time taken by the sound waves to travel from the solar surface to the Parker sonic critical point.

Introducing further,
\begin{equation}
    \label{eq::28''}
    u_1 \equiv \left( \epsilon \tau \right) v_1
\end{equation}
and  (\ref{eq::25}) again, equation (\ref{eq::27}) takes a form involving only the variables appropriate for the slowly-varying dynamics \tit{near} the Parker sonic criitcal point,
\begin{equation}
    \label{eq::29''}
    \frac{\del v_1}{\del \tilde{t}} + v_1 \frac{\del v_1}{\del x} = \left( \frac{x}{\tau}\right) v_1.
\end{equation}

Since equation (\ref{eq::29''}) is a nonlinear hyperboic equation, it is pertinent to use the method of characteristics (Garabedian 1986) to construct an analytical solution of this equation. Note that the characteristics $C$ of equation (\ref{eq::29''}) are given by
\begin{equation}
C: \hspace{.1in} \frac{d \tilde{t}}{d \xi} = 1, \hspace{.1in} \frac{dx}{d \xi} = v_1
\label{eq::29}
\end{equation}
Equation (\ref{eq::29''}) then reduces to the following ordinary differential equation,
\begin{equation}
\frac{d v_1}{d \xi} = \frac{1}{\tau} x(\xi)v_1(\xi), \hspace{.1in} \text{along C.}
\label{eq::30}
\end{equation}

Equation (\ref{eq::29}) yields the solution,
\begin{equation}
\tilde{t} = \xi, \hspace{.1in} x(x_0, \tilde{t}) = f(x_0, \tilde{t})
\label{eq::31}
\end{equation}
where,
$$
x_0 \equiv x(x_0,0).
$$
Using (\ref{eq::31}), equation (\ref{eq::30}) yields the solution,
\begin{equation}
v_1(x,\tau) = v_{1_0} e^{ \frac{1}{\tau} \int_0^{\tilde{t}} f(x_0, s)ds}.
\label{eq::32}
\end{equation}
where,
$$
v_{1_0} \equiv v_1(x,0).
$$

Introducing an auxilliary slow-time scale,
\begin{equation}
\psi(\tilde{t}) \equiv \frac{1}{\tau} \int_0^{\tilde{t}} f(x_0,s)ds
\label{eq::33}
\end{equation}
and using (\ref{eq::31}), we obtain
\begin{equation}
\frac{d \psi}{d \tilde{t}} = \frac{1}{\tau} f(x_0, \tilde{t}) = \frac{x}{\tau}
\label{eq::34}
\end{equation}
(\ref{eq::33}) and (\ref{eq::34}) imply the initial conditions,
\begin{equation}
\tilde{t} = 0: \hspace{.1in} \psi = 0, \hspace{.1in} \frac{d \psi}{d \tilde{t}} = \frac{x_0}{\tau}.
\label{eq::35}
\end{equation}

Furthermore, on using equations (\ref{eq::29}), (\ref{eq::31}) - (\ref{eq::34}), we have
\begin{equation}
\frac{d^2 \psi}{d \tilde{t}^2} = \frac{1}{\tau} \frac{dx}{d\tilde{t}} = \frac{v_{1}}{\tau} = \frac{v_{1_0}}{\tau} e^\psi
\label{eq::36}
\end{equation}
from which, we obtain
\begin{equation}
\frac{d \psi}{d \tilde{t}} = \sqrt{2 v_{1_0}/\tau}e^{\psi/2}.
\label{eq::37}
\end{equation}
We have from (\ref{eq::35}) and (\ref{eq::37}),
 \begin{equation}
 v_{1_0} = \frac{x_0^2}{2\tau}
 \label{eq::40}
 \end{equation}
On using (\ref{eq::40}), equation (\ref{eq::37}) yields the solution,
\begin{equation}
e^{-\psi/2} = 1-x_0 \frac{\tilde{t}}{2\tau}.
\label{eq::38}
\end{equation}
 Using (\ref{eq::38}), (\ref{eq::32}) and (\ref{eq::33}) give
 \begin{equation}
 v_1(x, \tilde{t}) = \frac{v_{1_0}}{(1-x_0 (\tilde{t}/2\tau ))^2}.
 \label{eq::39}
 \end{equation}
 Using (\ref{eq::40}), (\ref{eq::39}) becomes
 \begin{equation}
 v_1(x, \tilde{t}) = \frac{x_0^2/2\tau}{(1-x_0 (\tilde{t}/2\tau ))^2}.
 \label{eq::41}
 \end{equation}
 
 Using (\ref{eq::41}), equation (\ref{eq::29}) yields,
 \begin{equation}
 x(x_0, \tilde{t}) = \frac{x_0}{1-x_0 (\tilde{t}/2\tau )}
 \label{eq::42}
 \end{equation}
 from which, we obtain
 \begin{equation}
 x_0 = \frac{x}{1+x (\tilde{t}/2\tau )}.
 \label{eq::43}
 \end{equation}
 Using (\ref{eq::43}), (\ref{eq::41}) becomes
 \begin{equation}
 v_1(x,\tilde{t}) = \frac{1}{2\tau}x^2.
 \label{eq::44}
 \end{equation}
 (\ref{eq::28''}) then gives for the radial flow velocity perturbation,
 \begin{equation}
    u_1(x, \tilde{t}) = \frac{\epsilon}{2}x^2
    \label{eq::46}
 \end{equation}
 which is the exact solution of equation (\ref{eq::27})\footnote{
It may be mentioned that (\ref{eq::44}) turns out to belong also to a special set of solutions of equation (\ref{eq::29''}) given by the method of \emph{separation of variables}. Putting,
\begin{equation}
\tag{I}
v_1(x, \tilde{t}) = T(\tilde{t}) X(x)
\label{eq::I}
\end{equation}
equation (\ref{eq::27}) yields
\begin{equation}
\tag{II}
\frac{T'}{T} = (\frac{x}{\tau}-T X')
\label{eq::II}
\end{equation}
where prime denotes derivative with respect to the argument in question.

Equation (\ref{eq::II}) implies, on grounds of consistency,
\begin{equation}
\tag{III}
T(\tilde{t}) = \text{const.} = C
\label{eq::III}
\end{equation}
and hence,
\begin{equation}
\tag{IV}
X(x) = \frac{1}{2C\tau} x^2.
\label{eq::IV}
\end{equation}
Using (\ref{eq::III}) and (\ref{eq::IV}), (\ref{eq::I}) yields
\begin{equation}
\tag{V}
v_1(x, \tilde{t}) = \frac{1}{2\tau} x^2
\label{eq::V}
\end{equation}
in agreement with (\ref{eq::44}).
 }.Using (\ref{eq::46}), we obtain from (\ref{eq::25}) and (\ref{eq::26}), for the radial flow velocity perturbation,
 \begin{equation}
    \left( \frac{d \Phi}{dr} - a \right) = \epsilon a \frac{d \phi_1}{dr} = \frac{a}{2} \left(\frac{r}{r_*} - 1\right)^2, \hspace{.1in} \frac{r}{r_*} \approx 1.
    \label{eq::47}
 \end{equation}
 (\ref{eq::47}) implies that the leading dynamics in the nonlinear perturbation problem near the Parker sonic critical point is essentially frozen in time, and the perturbations remain stationary in time in this regime. Physically this seems to be traceable to the gravitational choking (described by the term on the right in equation (\ref{eq::27})) operational in the nonlinear hyperbolic dynamics near the Parker sonic critical point. Note that the spatial growth of the perturbations away from the Parker sonic critical point, as indicated by (\ref{eq::47}), does not become unbounded because this solution, as indicated by (\ref{eq::25}), is valid only near the Parker sonic critical point.\\

\section{Discussion.}

Contrary to the assumptions made in the theoretical models, the solar wind is, in reality, far from being steady and structureless, as revealed by spatial and temporal variabilities apparent in \textit{in situ} observations of the solar wind. Nonetheless, Parker's solar wind solution has been found to provide an excellent first-order approximation to the large-scale behavior, on the average, of the solar wind. This indicates it possesses a certain robustness and an ability to sustain itself against small perturbations acting on this system, hence rendering the stability of Parker's solar wind solution an important issue, though still not completely resolved. Previous investigations (Parker 1966, Carovillano and King 1966, Jockers 1968) of stability of Parker's solar wind solution with respect to linearized perturbations were hampered by the singularity at the \tit{Parker sonic critical point}, where the wind flow velocity equals the speed of sound in the wind. This paper seeks to regularize this singularity by going outside the framework of the linear perturbation problem, and incorporating the dominant nonlinearities in this dynamical system and hence provides an appropriate nonlinear recipe to regularize this singularity. This is implemented by introducing a whole new theoretical formulation, which is based on the \tit{potential flow theory} in ideal gas dynamics, and provides a smooth extrapolation of Parker's steady solar wind model to the unsteady case. The stability of Parker's solar wind solution is shown to extend also to the neighborhood of the Parker sonic critical point by going to the concomitant nonlinear problem. The new theoretical framework given here seems, therefore, to have the potential to provide a viable basis for future formulations addressing various theoretical aspects of Parker's unsteady solar wind model.
 
 It may be mentioned that deviations from isothermality in the solar wind may be described, in a first approximation, by using the polytropic gas model (Parker 1971, Holzer 1979, Shivamoggi and Pohl 2024),
 \begin{equation}
 p = C p^\gamma
 \label{eq::49}
 \end{equation}
 where $\gamma$ is the polytropic gas exponent, $1 < \gamma < 5/3$, and $C$ is an arbitrary constant. The generalization of the present formulations and results for polytropic solar wind flow will be reported in the near future. \\
 
 \centerline{*        *        *}
 
 This work was carried out during my sabbatical leave at California Institute of Technology. My thanks are due to Professor Shrinivas Kulkarni for his enormous hospitality and valuable discussions. I am thankful to late Professor Eugene Parker for his helpful advice and suggestions on the solar wind problem. I am thankful to Professors Earl Dowell, Grisha Falkovich, James Fuller, and Katepalli Sreenivasan for their helpful remarks and suggestions.

 \makeatletter
 \renewcommand\@biblabel[1]{}
 \makeatother

\end{document}